\documentclass[twocolumn,nofootinbib,amsmath,amssymb,prl,aps,superscriptaddress,tightenlines,preprintnumbers]{revtex4}

\usepackage{graphicx}
\usepackage{graphics}
\usepackage{braket}
\usepackage[usenames]{color}
 \usepackage[normalem]{ulem}
\DeclareMathAlphabet{\mathpzc}{OT1}{pzc}{m}{it}

\newcommand{\be}{\begin{equation}}
\newcommand{\ee}{\end{equation}}
\newcommand{\bea}{\begin{eqnarray}}
\newcommand{\eea}{\end{eqnarray}}
\newcommand{\ba}{\begin{array}}
\newcommand{\ea}{\end{array}}
\newcommand{\Lagr}{\mathcal{L}}

\newcommand{\nn}{\nonumber}

\def\nn{\nonumber\\ }

\def\hyp{\mathsf{y}}

\begin{document}

\newcount\hour \newcount\minute
\hour=\time \divide \hour by 60
\minute=\time
\count99=\hour \multiply \count99 by -60 \advance \minute by \count99
\newcommand{\mydate}{\ \today \ - \number\hour :00}

\title{The $ggh$ variations}

\author{
Adam Martin${}^{1}$ and Michael Trott${}^{1,2}$\\
${}^1$ Department of Physics, University of Notre Dame, Notre Dame, IN, 46556, USA \\
${}^2$ Niels Bohr Institute, University of Copenhagen,
Blegdamsvej 17, DK-2100, Copenhagen, Denmark}

\begin{abstract}
We examine how sub-leading results
in the operator and loop expansion
for $\sigma(\mathcal{G} \mathcal{G} \rightarrow h)$ in the Standard Model Effective Field Theory (SMEFT)
inform theoretical error estimates when studying this production channel in global SMEFT studies.
We also discuss the relationship between geometric SMEFT results and the $\kappa$ formalism.
 \end{abstract}

\maketitle
\newpage

\paragraph{\bf I. Introduction:}
The Standard Model (SM) is an incomplete description of observed phenomena in nature.
It must be extended to account for neutrino masses. This fact implies that new
physics will couple to the SM. In addition, the hierarchy problem also argues for an extended sector of new physics
at higher energy scales ($\Lambda$), if the origin of neutrino masses is associated
with such scales. As the exact origin of
neutrino masses and solution of the hierarchy problem is unknown, and certainly
experimentally unverified, it is useful to think of the SM as an Effective Field Theory (EFT)
for data analysis with characteristic energies around the electroweak scale:
$\bar{v}_T \equiv \sqrt{2 \, \langle H^\dagger H \rangle}$.

The Standard Model Effective Field Theory (SMEFT) is based on the low energy
assumptions that physics beyond the SM
is present at scales $\Lambda >\bar{v}_T$,
that there are no light hidden states in the spectrum with couplings
to the SM. A $\rm SU(2)_L$ scalar doublet ($H$) with Hyper-charge
$\hyp_h = 1/2$ is assumed present in the EFT. A power counting expansion in
the ratio of scales $q^2/\Lambda^2 <1$, follows with $q^2$ a kinematic invariant associated
with experimental measurements in the domain of validity of the EFT. These
low energy assumptions define a theory, the SMEFT, with a Lagrangian
\begin{align}
	\Lagr_{\textrm{SMEFT}} &= \Lagr_{\textrm{SM}} + \Lagr^{(5)}+\Lagr^{(6)} +
	\Lagr^{(7)} + \dots,  \\ \nonumber \Lagr^{(d)} &= \sum_i \frac{C_i^{(d)}}{\Lambda^{d-4}}\mathcal{Q}_i^{(d)}
	\quad \textrm{ for } d>4.
\end{align}
The operators $\mathcal{Q}_i^{(d)}$ are labelled with a mass dimension $d$ superscript
and multiply unknown Wilson coefficients $C_i^{(d)}$, that predict patterns of corrections to the SM.
The Wilson coefficients $C_i^{(d)}$ take on specific
values as a result of the $q^2/\Lambda^2 <1$ Taylor expanded effects of physics beyond the SM.
As the nature of physics beyond the SM is unknown, we treat the Wilson coefficients and $\Lambda$ as free parameters to fit
from the data, treating the SMEFT as its own --bottom up-- theory. For compact dimensionless notation we
define $\tilde{C}^{(d)}_i \equiv C^{(d)}_i \bar{v}_T^{d-4}/\Lambda^{d-4}$.
The SM Lagrangian notation and conventions are consistent with
Refs.~\cite{Buchmuller:1985jz,Grzadkowski:2010es,Alonso:2013hga,Brivio:2017vri,Brivio:2017btx,Helset:2020yio}.
The sum over $i$, after non-redundant operators are removed with field redefinitions
of the SM fields, runs over the operators in a particular operator basis.
We use the Warsaw basis \cite{Buchmuller:1985jz,Grzadkowski:2010es} for $\Lagr^{(6)}$ in this paper.

When projecting constraints
from global SMEFT fits onto the Wilson coefficients, one might expect to always use a theoretical prediction of the highest order in the operator
and the loop expansion. However, as one goes to higher order
in the SMEFT expansions, the number of unknown parameters continually increases
in a particular measurement increasing the resulting fit spaces.
This issue is not substantially ameliorated when combining multiple
measurements, as each measurement has this challenge of theoretical interpretation. It is necessary to
truncate the expressions to draw meaningful conclusions. The power counting of an EFT has a central
role as it organizes the infinite number of parameters that enter the predictions into sets
that are appropriate to retain when an approximate theoretical precision
(chosen to be better than the current experimental precision) is used to interface with the data.
For global SMEFT studies, the most straightforward choice is to retain all linear $\mathcal{L}^{(6)}$ interference terms
with the SM amplitudes.

However, for $\sigma(\mathcal G\mathcal G \to h)$, this choice faces challenges. Because the SM amplitude itself is loop level, parts of the SMEFT calculation that interfere with the SM -- such as the piece linear in $(\mathcal{L}^{(6)}$)  -- are suppressed compared to quadratic terms -- $(\mathcal{L}^{(6)})^2$. The quadratic term is accompanied by a factor of $(\bar v_T/\Lambda)^2$ relative to the interference piece, but for low $\Lambda$ this may not be enough to compensate for the loop factor difference. This interplay of loop factors and $\bar v_T/\Lambda$ can be further exacerbated if one assumes hierarchical Wilson coefficients, as has been shown to arise in many UV matching scenarios \cite{Arzt:1994gp, deBlas:2017xtg, Craig:2019wmo} and, more generally, follows from the conditions of naive ($d \leq 4$) renormalizability being imposed on all UV physics at higher scales \cite{Jenkins:2013fya}. Specifically, if the relevant dimension six coefficients are small as the result of UV matching, while the dimension eight coefficients are order one, both the $\mathcal{L}^{(6)}$ and $(\mathcal{L}^{(6)})^2$ contributions to $\sigma(\mathcal G\mathcal G \to h)$ may be subdominant to $(\mathcal L^{(8)})$ terms. This argument applies to all SM loop processes, however we will focus on $\sigma(\mathcal G\mathcal G \to h)$ given the prominent role it plays in SMEFT global fits. An analysis of $\Gamma(h \to \gamma\gamma)$ is given in Appendix {\bf B}.

In this paper, we explore how choices about where the SMEFT calculation is truncated when interpreting experimental results and what is assumed about the hierarchy among Wilson coefficients affect the theoretical error on $\sigma(\mathcal G\mathcal G \to h)$. There is no unique answer to defining an error estimate for neglected higher order terms in a perturbative expansion, and a reasonable error estimate is never an assertion of precise and exact knowledge of all higher order terms.\footnote{Nevertheless such errors are still meaningful and standard to incorporate in EFT studies for decades.}
Here we restrict ourself to
a well defined procedure for defining such an error,
maximally informed by the actual higher order results, when such results are available
in the literature. For $\sigma(\mathcal G\mathcal G \to h)$, the result including both $\mathcal L^{(6)}$ effects to one loop order  -- $\mathcal O(1/16\pi^2\Lambda^2)$ -- and the complete set of $\mathcal L^{(8)}$ effects  -- $\mathcal O(1/\Lambda^4)$ -- was recently developed in Ref.~\cite{Corbett:2021cil}.\\

\paragraph{\bf II. $\sigma(\mathcal{G} \mathcal{G} \rightarrow h)$ to $\mathcal O(1/\Lambda^4), \mathcal O(1/16\pi^2\Lambda^2)$:}
It is appropriate to organize $\Lagr_{\textrm{SMEFT}}$ as specific composite operator
kinematics, with scalar dressings that do not introduce new kinematics, to identify the full set of $\mathcal{O}(1/\Lambda^4)$
corrections. This is the geoSMEFT approach developed
in Refs.~\cite{Helset:2018fgq,Helset:2020yio,Hays:2020scx,Corbett:2020bqv}
where scalar field dependent field-space connections $G_i$
multiply composite operator forms $f_i$ as
\bea\label{basicdecomposition}
\Lagr_{\textrm{SMEFT}} = \sum_i G_i(I,A,\phi \dots) \, f_i.
\eea
Powers of $D^\mu H$ are included in $f_i$, $I$ and $A$ represent possible ${\rm SU(2)_W}$ and ${\rm SU(3)}$ group structures, and $\phi_{1,2,3,4}$ are components of the
Higgs $H$ field.
The kinematic dependence is factorized into the $f_i$ and the re-scalings by $G_i$.
The geoSMEFT is defined to all orders in the $\sqrt{2 \langle H^\dagger H\rangle/\Lambda}$
expansion for low $n$-point functions ($n \leq 3$), which is sufficient for the case of interest
here. In addition, as the loop expansion and
the operator expansion are not independent at sub-leading order in the SMEFT \cite{Corbett:2021cil},
it is necessary to formulate $\mathcal{O}(1/16 \pi^2 \Lambda^2)$ corrections in the SMEFT in a manner consistent with the geoSMEFT
organization higher order $\mathcal{O}(1/\Lambda^4)$ physics. This is best accomplished in the Background Field Method (BFM) approach
to gauge fixing in the SMEFT \cite{DeWitt:1967ub,tHooft:1973bhk,Abbott:1981ke,Helset:2018fgq}.

The general Higgs-gluon field space metric is defined as \cite{Helset:2020yio}
\bea
\Lagr_{\textrm{SMEFT}} \supset - \frac{1}{4}  \kappa(\phi) G^{\mathpzc{A},\mu \nu} G_{\mathpzc{A},\mu \nu},
\label{eq:kappadef}
\eea
with $\mathpzc{A}$ running over $\{1 \cdots 8 \}$ and
\bea
\kappa(\phi)= \left(1- 4 \, \sum_{n=0}^\infty C_{HG}^{(6+ 2n)} \, \left(\frac{\phi^2}{2}\right)^{n+1} \right).
\eea
For the gluon field strength and coupling, the transformations to canonically normalized fields at all $1/\Lambda^n$ orders are given by
\begin{align}\label{allordersnorm}
G^{A,\nu} &=  \sqrt{\kappa} \, \mathcal{G}^{\mathpzc{A},\nu}, \\
\bar{g}_3 &= g_3 \, \sqrt{\kappa}.
\end{align}
We return to the nature of these field redefinitions below.

We write the amplitude perturbation to the process as \cite{Corbett:2021cil}
\begin{align}
\mathcal \mathcal{A}_{\mathcal{G} \mathcal{G}h} &= \mathcal \mathcal{A}^{\mathcal{G} \mathcal{G}h}_{SM} + \langle \mathcal{G} \mathcal{G}|h \rangle^0_{\mathcal O(\bar{v}_T^2/\Lambda^2)} + \langle \mathcal{G} \mathcal{G}|h \rangle^1_{\mathcal O(\bar{v}_T^2/\Lambda^2)}, \nn
&+  \langle \mathcal{G} \mathcal{G}|h \rangle^0_{\mathcal O(\bar{v}_T^4/\Lambda^4)} + \cdots
\end{align}
where each of the expressions for $\mathcal \mathcal{A}^{\mathcal{G} \mathcal{G}h}_{SM}$, $\langle \mathcal{G} \mathcal{G}|h \rangle^0_{\mathcal O(\bar{v}_T^2/\Lambda^2)}$
$ \langle \mathcal{G} \mathcal{G}|h \rangle^1_{\mathcal O(\bar{v}_T^2/\Lambda^2)}$ and $\langle \mathcal{G} \mathcal{G}|h \rangle^0_{\mathcal O(\bar{v}_T^4/\Lambda^4)}$
are now known in a consistent set of perturbations in the loop (indicated with a super-script number) and operator expansion (indicated with a sub-script).
The SM result itself $\mathcal \mathcal{A}^{\mathcal{G} \mathcal{G}h}_{SM}$ also has a perturbative expansion,
and is often determined in an operator expansion with a heavy top limit taken. Here we use the SM result
as reported in Ref.~\cite{Corbett:2021cil}, which is not the highest order SM result known, but sufficient for our
error estimate purposes. The important aspects of the series behavior is the interaction of
new parameters appearing perturbing the SM, and the appearance of perturbative loop correction factors $\propto 1/16 \pi^2$
and operator expansion corrections $\propto \bar{v}_T^2/\Lambda^2$.

The operator expansion of the field space connection
introduces sensitivity to one new $\mathcal{L}^{(8)}$ operator at subleading order: $\mathcal{Q}^{(8)}_{HG}$.
In addition, several cross terms of $\mathcal{L}^{(6)} \times \mathcal{L}^{(6)}$ form,
including an important contribution from $(C^{(6)}_{HG})^2$ are present in the expansion of $\sqrt{\kappa}$. Similarly, the SMEFT loop expansion introduces corrections to the Wilson coefficient already present at leading order
($C_{HG}^{(6)}$) and also introduces the new parameters ($C_{H \Box}^{(6)}$, $C_{HD}^{(6)}$, $C_{Hu}^{(6)}$, $C_{uG}^{(6)}$)
whose operator definitions are
\begin{align}
\mathcal{Q}_{H \Box}^{(6)} &= (H^\dag H)\Box(H^\dag H), \nn
\mathcal{Q}_{HD}^{(6)} &= \ \left(H^\dag D_\mu H\right)^* \left(H^\dag D_\mu H\right), \nn
\mathcal{Q}_{Hu}^{(6)} &= (H^\dag i\overleftrightarrow{D}_\mu H)(\bar u_r \gamma^\mu u_r), \nn
\mathcal{Q}^{(6)}_{uG} &= (\bar q_3 \sigma^{\mu\nu} T^A u_3) \widetilde H \, G_{\mu\nu}^A,
\end{align}
here $r,s$ run over $1,2,3$ for the up ($u$), charm ($c$) and top ($t$) quark flavor labels
and $\widetilde H_j = \epsilon_{jk} \, H^{\dagger,k}$.
For remaining notational conventions consult Refs.~\cite{Buchmuller:1985jz,Grzadkowski:2010es,Alonso:2013hga,Brivio:2017vri,Brivio:2017btx,Helset:2020yio}.
Dependence on
\begin{align}
\delta G_F^{(6)} &= \frac{1}{\sqrt2} \left(\tilde C^{(3)}_{\substack{Hl \\ee}}+\tilde C^{(3)}_{\substack{Hl \\ \mu \mu}} - \frac{1}{2}(\tilde C'_{\substack{ll \\ \mu ee \mu}}+\tilde C'_{\substack{ll \\ e \mu \mu e}})\right)
\end{align}
is also present due to a redefinition of the input parameter vev, introducing a further dependence on the coefficients of
\begin{align}
\mathcal{Q}^{(3)}_{\substack{Hl \\pr}} &=(H^\dag i\overleftrightarrow{D}^I_\mu H)(\bar l_p \tau^I \gamma^\mu l_r),\nn
\mathcal{Q}'_{\substack{ll \\prrp}} &= (\bar l_p \gamma^\mu l_r)(\bar l_r \gamma^\mu l_p)
\end{align}
The explicit expression defined/developed in
Ref.~\cite{Corbett:2021cil} is
\begin{widetext}
\begin{align}
\frac{ \sigma^{\hat{\alpha}}_{\rm SMEFT}(\mathcal G\mathcal G \to h)}{\sigma^{\hat{\alpha}, 1/m^2_t}_{\rm SM}(\mathcal G\mathcal G \to h)}\simeq 1 & +  519\, \tilde C^{(6)}_{HG} + 504\, \tilde C^{(6)}_{HG}\Big(\tilde C^{(6)}_{H\Box} - \frac 1 4 \tilde C^{(6)}_{HD} \Big) + 8.15\times10^4\, (\tilde C^{(6)}_{HG})^2 + 504\, \tilde C^{(8)}_{HG} \nn
&\hspace{-2cm} + 1.58\, \Big(\tilde C^{(6)}_{H\Box} - \frac 1 4 \tilde C^{(6)}_{HD} \Big) + 362\, \tilde C^{(6)}_{HG} -1.59\, \tilde C^{(6)}_{uH} - 12.6\, {\rm Re }\, \tilde C^{(6)}_{uG}
 - 1.12\,\delta G^{(6)}_F - 7.70\,  {\rm Re }\, \tilde  C^{(6)}_{uG}\,\log\Big(\frac{\hat m^2_h}{\Lambda^2} \Big).
\label{eq:gghrationumeric}
\end{align}
\end{widetext}
The expression is reported in the $\hat{\alpha}_{ew}$ input parameter scheme, but input parameter scheme dependence is
negligible in this expression. In this expression, we have omitted contributions from Yukawa couplings other than $y_t$ as they are numerically negligible~\footnote{We also ignore all CP odd operators due to strong, low energy constraints, see Refs.~\cite{Cirigliano:2016njn,Cirigliano:2016nyn}}. The leading dependence on $\tilde C^{(6)}_{HG}$  has a numerical coefficient $519$, this coefficient is a few percent different than the coefficient of $\tilde C^{(8)}_{HG}$ as in $\mathcal \mathcal{A}^{\mathcal{G} \mathcal{G}h}_{SM}$
we have expanded in the heavy top limit and retained a higher order term in the former.

To explore the effect of retaining higher order terms in the
interpretation of a projected measurement of $\sigma(\mathcal{G} \mathcal{G} \rightarrow h)$
from the production and decay of the Higgs with theory errors, we consider three cases. In each case, the full expression in Eq.~\eqref{eq:gghrationumeric} is broken up into a piece used to project experimental constraints, and the remainder, which represents neglected higher order terms. The three cases are:
\begin{itemize}
\item[i.)]  Interpret experimental data using the linear $\mathcal L^{(6)}$ interference term only, which in this case is just the $C_{HG}^{(6)}$ contribution. We include both the tree level and one loop correction $\propto C^{(6)}_{HG}$.
\item[ii.)] Interpret experimental data keeping the $C_{HG}^{(6)}$ interference term plus the $(C_{HG}^{(6)})^2$ `squared' piece.
\item[iii.)] In addition to the pieces in ii.), retain the $C^{(8)}_{HG}$ contribution.
\end{itemize}

While case i.) is the standard, the validity and features of cases ii.) and iii.) warrants more study before jumping into numerics. \\

\paragraph{\bf III. Quadratic fits in Loop processes and Tree level processes in the SM:}
Retaining terms in the SMEFT prediction for the practical purpose of theoretical precision
being greater than experimental precision would naively argue for retaining the $(C_{HG}^{(6)})^2$ term
unless $C_{HG}^{(6)} \ll 1$, due to (for example) a loop suppression in matching.

On the other hand, retaining only a subset of terms at an order in a power counting expansion
is ill-defined in EFT formally, as a field redefinition on a SM field $F$
\bea
F \rightarrow F'[1 + \mathcal{O}(1/\Lambda^n)]
\eea
can change the set of parameters retained (or remove the parameters entirely), resulting in ambiguous predictions. This point was recently stressed in Ref.~\cite{Trott:2021vqa}.

For example, a field redefinition involving $C_{HG}^{(6)}$ is allowed (and even required on the gluon field to take the theory to canonical form)
in Eq.~\eqref{allordersnorm} at $\mathcal{O}(1/\Lambda^4)$.  This field redefinition on $\mathcal \mathcal{A}^{\mathcal{G} \mathcal{G}h}_{SM}$ cancels
order by order against the simultaneous redefinition of the gauge coupling at all orders (see Eq.~\eqref{allordersnorm}).
Applied to $\mathcal{Q}_{HG}$, the redefinition
\begin{align}
G^A_{\mu\nu} \to G^A_{\mu\nu}( 1 + O(C^{(6)}_{HG}\,v^4_T/\Lambda^4))
\end{align}
does not cancel and generates $\mathcal O ((C^{(6)}_{HG})^2/\Lambda^4)$ effects that are ambiguous until the theory is fully defined at $\mathcal{O}(1/\Lambda^4)$. However, when determining the cross section, the ambiguous $\mathcal O ((C^{(6)}_{HG})^2/\Lambda^4)$ terms enter via interference with the  (loop suppressed) SM amplitude $\mathcal \mathcal{A}^{\mathcal{G} \mathcal{G}h}_{SM}$ and are therefore numerically suppressed (regardless of how one chooses the Wilson coefficients) compared to the quadratic (self-square) contribution
-- $(C^{(6)}_{HG})^2$. In this sense, quadratic fits to {\it loop suppressed processes in the SM}, although formally
inconsistent in the treatment of the power counting, are only sensitive to a small numerical error/ambiguity in some cases.
This is the case when considering the Higgs-Gluon field space connection, and constraints on $\mathcal{Q}_{HG}^{(6)}$
retaining quadratic terms are of increased interest as a result.


This reasoning only applies to $\mathcal{Q}^{(6)}_{HG}$, $(\mathcal{Q}_{HG}^{(6)})^2$ when studying constraints on $\sigma(\mathcal{G} \mathcal{G} \rightarrow h)$ and fails -- in the sense that relative numerical errors are subsequently $\mathcal{O}(1)$ -- for all other Wilson coefficent dependence
in Eq.~\eqref{eq:gghrationumeric}. In particular, it fails for dependence on the Wilson coefficient of $\mathcal{Q}^{(8)}_{HG}$. \\

\paragraph{\bf IV. $\kappa$ rescalings and geoSMEFT}
It is interesting to consider the
possibility of projecting experimental constraints on the entire Higgs-Gluon field space connection $\kappa$ (defined in Eq.~\eqref{eq:kappadef}),
and this relationship of such a procedure to the so called ``$\kappa$ formalism"
developed in Refs.~\cite{Duhrssen:2004cv,Espinosa:2012ir,Carmi:2012yp,Azatov:2012bz,LHCHiggsCrossSectionWorkingGroup:2012nn}.\footnote{The coincidence
in notion should not be over interpreted.}

In the ``$\kappa$ formalism", the coefficient of the three point $g$-$g$-$h$ coupling is treated as parameter that experiments fit to. We can map the geoSMEFT expression, Eq.~\eqref{eq:kappadef} into this form by expanding to linear order in $\phi$
\begin{align}
\kappa_{\rm geoSMEFT} = \langle \frac{\delta \kappa}{\delta h} \rangle \langle \kappa \rangle =
- 4 \, \frac{\tilde{C}^{(6)}_{HG}}{\bar{v}_T}
- 4 \, \frac{\tilde{C}^{(8)}_{HG}}{\bar{v}_T}
+ 8 \, \frac{(\tilde{C}^{(6)}_{HG})^2}{\bar{v}_T}.
\end{align}
One may expect that $\kappa_{\rm geoSMEFT}$ is less sensitive to Wilson coefficient hierarchies, such as the tree/loop scenario, where $C^{(8)}_{HG} \sim 16 \pi^2 C^{(6)}_{HG} /g^2$, since all effects are lumped into a single coefficient. However, when inspecting the cross section ratio (Eq.~\eqref{eq:gghrationumeric}), $\kappa_{\rm geoSMEFT}$ is not manifest. Treating the $g$-$g$-$h$ vertex as a single object misses subtleties, such as which terms interfere with the SM and which do not, that the operator expansion catches.

Extracting the components of Eq.~\eqref{eq:gghrationumeric} that have the largest numerical factors and fewest powers of the $\tilde C_i$, we find some middle ground -- a quantity that involves only a few Wilson coefficients yet is derived at the cross section level so captures information about interference with the SM.
\begin{align}
&\frac{ \sigma^{\hat{\alpha}}_{\rm SMEFT}(\mathcal G\mathcal G \to h)}{\sigma^{\hat{\alpha}, 1/m^2_t}_{\rm SM}(\mathcal G\mathcal G \to h)} \simeq 1 + 881\,\Sigma_k  + \cdots \nonumber \\
&\Sigma_\kappa =  \left[\tilde C^{(6)}_{HG} + 0.57 \, \tilde C^{(8)}_{HG} + 93 \,(\tilde C^{(6)}_{HG})^2\right]
\end{align}
The  coefficient
$881$ is the sum of the tree level  $\tilde C^{(6)}_{HG}$ term plus the retained loop correction
for this operator; the relative $0.57$ in front of  $\tilde C^{(8)}_{HG}$ comes about because terms of $\mathcal{O}(\bar{v}_T^4/16 \pi^2 \Lambda^4)$ were not included in Ref.~\cite{Corbett:2021cil}. Had these terms been included the factor $0.57 \rightarrow \sim 1$.
Fitting to $\Sigma_k$ corresponds to case iii).

The combination $\Sigma_k$ is present for other phenomena involving a single Higgs. For example,
the significant numerical dependence on $\tilde{C}^{(6)}_{HG}$ in the Higgs width in the SMEFT \cite{Brivio:2019myy}
can be rescaled out using the results in \cite{Corbett:2021cil} as
\bea\label{widthrescale}
\frac{\Gamma_{h,full}^{SMEFT}}{\Gamma_h^{SM}}
&\simeq& \,  1
+ 50.6    \,\tilde{C}^{(6)}_{HG} + \cdots \\
&\simeq& 1+  88  \Sigma_\kappa - 6.7 \tilde{C}^{(6)}_{HG} + \cdots \nonumber
\eea
The total Higgs width has a very significant dependence on $(\tilde{C}^{(6)}_{HG})^2$
in the SMEFT via $\Sigma_\kappa$. The subtraction of an explicit dependence on $\tilde{C}^{(6)}_{HG}$
is due to  the difference in the one loop correction in $\sigma(\mathcal{G}\mathcal{G} \rightarrow h)$ vs
$\Gamma(h\rightarrow \mathcal{G}\mathcal{G})$ at one loop as specified in Ref.~\cite{Corbett:2021cil}.

To break the parameter degeneracy built into $\Sigma_\kappa$ experimentally one needs a
consider a process with more than one Higgs field exchange at tree level in a Feynman diagram, or further
loop corrections. For example, the parameter degeneracy in of  $\tilde{C}^{(6)}_{HG}$ in
$\sigma(\mathcal{G}\mathcal{G} \rightarrow h)$ and
$\Gamma(h\rightarrow \mathcal{G}\mathcal{G})$ is already weakly broken by a one loop correction,
as shown in Eq.~\eqref{widthrescale}.

In general, the geoSMEFT approach is closely related to the $\kappa$ formalism
where rescalings of SM processes occurs with common kinematic dependence in the SM
and an effective field theory extension. It has been argued that the specific implementation
of this idea in Ref.~\cite{LHCHiggsCrossSectionWorkingGroup:2012nn} is directly mappable
to the HEFT formalism in Ref.~\cite{Buchalla:2015wfa, Buchalla:2015qju}. The geoSMEFT
also provides a rescaling generalization of the SM which allows a field theory interpretation
of the $\kappa$ formalism in Ref.~\cite{LHCHiggsCrossSectionWorkingGroup:2012nn},
that can also be extended to non-SM kinematics in a well defined way. The resummation of higher
orders in $\bar{v}_T^2/\Lambda^2$ in the geometric dressings of the composite operator
forms also breaks the relationships between SMEFT corrections enforced by linearly realized
${\rm SU(2)_L}$ symmetry, as in the HEFT. However, in the geoSMEFT case the expansion
back to a linear realization SMEFT is direct and follows from Taylor expanding
the geoSMEFT field space connections.\\

\paragraph{\bf V. Numerical study} To more quantitatively understand the impact of including higher order terms in the interpretation of experimental $\sigma(\mathcal G \mathcal G \to h)$ data, we turn to numerics. Specifically, we study how the uncertainty -- encapsulated by the remainder terms for the cases identified earlier -- varies among the cases and as we change assumptions about the sizes of Wilson coefficients. We consider two different Wilson coefficient schemes, a) all Wilson coefficients set to the same value, and
b) an ordering of the Wilson coefficients according to a tree-loop matching scheme. Defining the uncertainty in this fashion is consistent with the
arguments in Refs.~\cite{Hays:2020scx,Corbett:2021eux,Corbett:2021cil} and in particular Ref.~\cite{Trott:2021vqa}.

\begin{figure}[h!]
\centering
\includegraphics[width=0.45\textwidth]{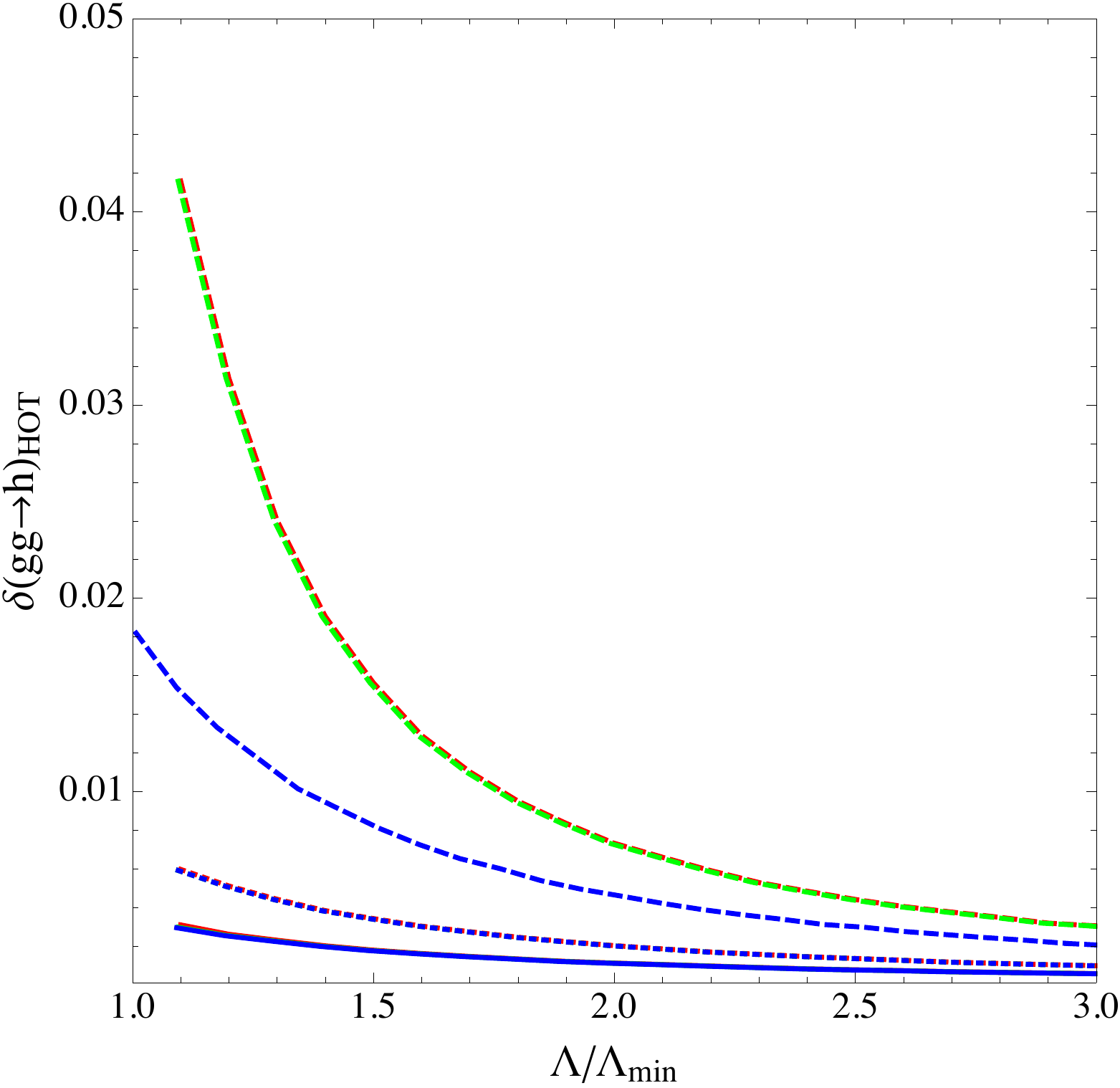}
\caption{Uncertainty on $\sigma(gg \to h)$ from higher order terms
 as a function of the new physics scale $\Lambda$ relative to the minimum scale compatible with current
 experimental $gg \to h$ data. We have broken the full $\sigma(gg \to h)_{SMEFT}$ calculation of
 Ref.~\cite{Corbett:2021cil} into a calculation piece used to determine the
 compatibility scale -- and a higher order terms piece in three different ways,
 i.) first we retain only the $\mathcal O(1/\Lambda^2)$ interference term,
 ii.) second we include the interference term and pieces proportional to $(C^{(6)}_{HG})^2$
 in the retained result in the calculation to determine the compatability scale, iii.)
 as in ii.) but also the $C^{(8)}_{HG}$ term is also included, which corresponds to $\Sigma_\kappa$.
 For a given set of terms retained, we determine the minimum scale by equating it with the current experimental
 uncertainty on $\sigma(\mathcal G \mathcal G \to h)$. The curves shown above are then
 generated by incrementing $\Lambda$ above $\Lambda_{\rm min}$ and numerically
 evaluating the numerical error by plugging in coefficients according to a scheme and evaluating the neglected terms.
 The dashed lines correspond to the tree/loop scheme with values 1.0/0.01, the solid
 lines correspond to picking all coefficients equal to 0.01, and the dotted lines correspond to picking all coefficients equal to 1.0.
Case i.) is shown in red, case ii.) in green, case iii.) in blue. For the tree/loop scheme, the retained partitions i.) and ii.)
 are nearly identical, while all three cases are nearly identical when all coefficients are
 chosen equal (same color scheme for the cases).
\label{fig:hoterror}}
\end{figure}

Our first step is to focus our study on coefficients and $\Lambda$ scales that are not already experimentally excluded. We do this by equating the retained piece of the SMEFT calculation in each case to  the current experimental
uncertainty on $\mu_{ggh}$, e.g. for case i.) we solve $881\, \tilde C^{(6)}_{HG}  = \delta\mu_{ggh}$. To extract a rough minimum $\Lambda_{\rm min}$ scale from this, we plug in for $C^{(6)}_{HG}$ according to the Wilson coefficient scheme. We take this constraint from a fit to $\mu_{ggh}$ taking the constraint $\mu_{ggh} = 1.04 \pm 0.09$~\cite{Aad_2020, CMS-PAS-HIG-19-005}, using $0.09$ as a rough error band to define relevant perturbations that are not experimentally disfavored when considering error estimates.\footnote{Note that significant cancellations can occur between terms in Eqn.\eqref{eq:gghrationumeric}, lowering a naive compatability
scale, and this translates into cases where the theory error on the experimental projection of results onto $C_{HG}^{(6)}$ etc is significantly higher.
This is also qualitatively indicated with the blowing up of the theory error curve in Fig.~\ref{fig:hoterror}.
Such cancellations, leading to flat directions, are broken by considering top measurements \cite{Degrande:2012gr}
and in a global study are expected to be less relevant than the generic case considered here with a naive compatability scale
and no significant cancellations. Such potential cancellations, with a corresponding large theory error when canceling terms are neglected, are also illustrated
by the variations in Figs.~\ref{fig:all01},\ref{fig:all10},\ref{fig:treeloop}.}

Next, we numerically evaluate the uncertainty for the three cases as a function of $\Lambda > \Lambda_{\rm min}$. To avoid accidental cancellations, we assign values to the Wilson coefficients at each step by drawing them from gaussian distributions centered at zero and with widths set by the Wilson coefficient scheme. Repeating this $10{\rm k}$ times at each $\Lambda$ step, we take the $1\sigma$ width of the resulting gaussian distribution as the theory error. This theory error is driven primarily by $\Lambda$, and is, by design, restricted to scales that are still viable for a given coefficient choice. Switching to a flat distribution for sampling the Wilson coefficients leads to identical results. This is to be expected; evaluating the uncertainty in this way amounts to sampling the linear sum of multiple parameters, so the central limit theory dictates that the resulting error distribution will be gaussian regardless of how the individual terms are sampled.

The resulting theory error is shown in Fig.~\ref{fig:hoterror}. For the tree/loop Wilson coefficient matching scheme we use values of 1.0/0.01. For the matching scheme with all coefficients taken equal, we try two values, all coefficients 0.01 and all 1.0. The $\Lambda_{\rm min}$ values for the cases are different\footnote{Explicitly, for all coefficients equal to 0.01, $\Lambda_{\rm min} = 2.43\, {\rm TeV}, 2.44\, {\rm TeV}, 2.45\, {\rm TeV}$ for cases i.), ii.) iii.) respectively, for all coefficients equal to 1.0, $\Lambda_{\rm min} = 24.3\, {\rm TeV}, 24.5\, {\rm TeV}, 24.5\, {\rm TeV}$, and $\Lambda_{\rm min} = 2.43\, {\rm TeV}, 2.44\, {\rm TeV}, 2.90\, {\rm TeV}$ for the tree/loop (1.0/0.01) scheme.}, but plotting the curves versus $\Lambda/\Lambda_{\rm min}$ hides shifts in $\Lambda_{\rm min}$ and allows all curves to be shown on one plot. The solid (all coefficients 0.01)  and dotted (all coefficients 1.0) lines are nearly identical, as overall changes in the coefficients can be compensated -- up to the terms containing $\log(\Lambda^2)$ -- by rescaling  $\Lambda_{\rm min}$.

 When all coefficients are chosen equal, the error estimate in all the cases is nearly identical. When coefficients are chosen with the tree/loop hierarchy, the error is case iii.) is roughly two times smaller than cases i.), ii.). This difference is due to $C^{(8)}_{HG}$, a tree level term as classified by Ref.~\cite{Craig:2019wmo}, that is part of the uncertainty in cases i.) and ii.) but not in case iii.). The size and stability of the uncertainty curve for case iii.) under the two Wilson coefficient matching schemes makes the case for projecting experimental fit results onto $\Sigma_\kappa$. An alternative theory error analysis, fixing $C^{(6)}_{HG}$ and sampling the higher order terms using the method of Ref.~\cite{Hays:2020scx}, is shown in Appendix {\bf A}.

As $\Sigma_\kappa$ is not the complete $\mathcal O(1/\Lambda^4)$ result, the obvious worry is that there may be artifacts or ambiguities present. The analysis of Sec. {\bf III} shows that field redefinition ambiguities can be present, but are small. A second concern is that  combination of parameters retained introduces intrinsic basis choice dependence. For example, it has been shown in Refs.~\cite{Hays:2020scx,Corbett:2021eux} that dependence on
$\lambda \bar{v}_T^2$ purely due to operator basis choice in matching a UV model onto the SMEFT cancels
in observables, but could persist in inconsistent calculations to $\mathcal{O}(1/\Lambda^4)$.
To check whether or not this combination
of terms introduces such an intrinsic basis dependence, we study a matching example.
\\

\paragraph{\bf VI. Matching example}
Consider integrating out $\sigma$, which couples to the SM as
\begin{align}
\mathcal L_\sigma = \frac 1 2 (\partial_\mu\sigma)^2 - \frac 1 2 m^2_\sigma \sigma^2 + \frac {a g^2_3\, \sigma G^A_{\mu\nu}G^{A, \mu\nu}}{\Lambda} + \Lambda\, b\, H^{\dag}H\sigma
\end{align}
This is an example of a non-minmally coupled model, as discussed in Ref.~\cite{Jenkins:2013fya}.
Rewriting the Lagrangian as
\bea
\mathcal L &=& \mathcal L_{SM} - \frac 1 2 \sigma (\partial^2 + m^2_{\sigma} )\sigma + \sigma \, B, \\
B &=& \frac {a g^2_3\ G^A_{\mu\nu}G^{A, \mu\nu}}{\Lambda} + \Lambda\, d\, H^{\dag}H
\eea
then using the results from Ref.~\cite{Henning:2014wua,Corbett:2021eux} yields:
\begin{align}
\mathcal L = \mathcal L_{SM} + \frac 1 {2m^2_\sigma}B^2 + \frac 1 {2m^4_\sigma}B \partial^2 B + \mathcal O(m^{-6}_\sigma).
\end{align}
It is necessary for the condition $b \ll \Lambda/m_\sigma$ to be imposed for the expansion in $1/m_\sigma$ to be convergent.
The low energy effects of this matching is to redefine the SM $\lambda$, $\bar{v}_T$ and $m_h$ as
\bea
\bar{v}_T^2 &\rightarrow & (v')^2  + \frac{b^2 \, \Lambda^2 \bar{v}_T^2}{2 \,\lambda \,m_\sigma^2}, \\ \lambda &\rightarrow& \lambda' - \frac{b^2 \, \Lambda^2}{2 m_\sigma^2}, \\
m_h^2 &=& 2 \, \lambda \bar{v}_T^2, \\
 &\rightarrow& 2 \, \lambda' \, (v')^2 \left[1- \frac{b^4 \Lambda^4}{4 (\lambda')^2 \, m_\sigma^4} \right].
\eea
The remaining contributions come from expanding out $B \partial^2 B$. The effects include a contribution to the gluon self interactions
\bea
\frac{1}{2 m_\sigma^4} \frac{a^2 \, g_s^4}{\Lambda^2} \left[G^A_{\mu\nu}G^{A, \mu\nu} \right]\partial^2 \left[G^A_{\mu\nu}G^{A, \mu\nu}\right],
\eea
and a contribution to $C_{H \Box}$ is
\bea
\frac{\Lambda^2}{2 m_\sigma^4} b^2 \mathcal{Q}_{H\Box}.
\eea
A more interesting interaction comes from the cross term
\bea\label{derivativeterms}
\frac{a \, b \, g_s^2}{2 m_\sigma^4} \, \left[G^A_{\mu\nu}G^{A, \mu\nu} \, \partial^2 (H^\dagger H)
+  \partial^2 (G^A_{\mu\nu}G^{A, \mu\nu}) \, (H^\dagger H)\right]. \nonumber
\eea
One can integrate by parts to arrange this contribution into the form
\bea
\frac{a \, b \, g_s^2}{2 m_\sigma^4} \, \left[2 G^A_{\mu\nu}G^{A, \mu\nu} \, \partial^2 (H^\dagger H)\right].
\eea
Expanding out
\bea
\partial^2 (H^\dagger H) \rightarrow 2 (D^\mu H^\dagger) (D_\mu H)+ (D^2 H^\dagger) H + H^\dagger (D^2 H). \nonumber
\eea
The first term does not contribute to $\sigma(gg \rightarrow h)$, the remaining terms are EOM reducible to the combination of terms
\bea
2 \left(\lambda \bar{v}_T^2 (H^\dagger H) - 2 \lambda (H^\dagger H)^2 + {\rm Yukawa \, terms} \right)
\eea
Combining these with the $O(m^{-2}_\sigma)$ term, we have
\begin{align}
\mathcal L \supset \Big( \frac {a b\, g^2_3}{m^2_\sigma} + \frac{2a\, b\, g^2_3\, \lambda v^2}{m^4_\sigma}\Big) \mathcal Q^{(6)}_{HG} - \frac{4a\, d\, g^2_3\, \lambda}{m^4_\sigma} \mathcal Q^{(8)}_{HG}
\end{align}
Via the expression $\Sigma_\kappa$, this leads to a $\lambda$ dependence in the cross section
\bea
\frac{ \sigma^{\hat{\alpha}}_{\rm SMEFT}(\mathcal G\mathcal G \to h)}{\sigma^{\hat{\alpha}, 1/m^2_t}_{\rm SM}(\mathcal G\mathcal G \to h)} \propto - 1.7 \times 10^3 \frac{a \, b \, \lambda \, \bar{v}_T^4}{m_\sigma^4}
\eea
when we incorporate the common one loop QCD correction to $\mathcal{Q}_{HG}^{(8)}$ to have a common tree level
dependence. This arrangement of derivative terms
is consistent with the geoSMEFT conventions.
However, unlike the examples in Refs.~\cite{Hays:2020scx,Corbett:2021eux} this $\lambda$ dependence
does not signal intrinsic basis dependence in fitting to $\Sigma_\kappa$ due to an inconsistent
treatment of the theory at $\mathcal{O}(1/\Lambda^4)$. One can also rearrange the derivative terms onto
\bea
\frac{a \, b \, g_s^2}{2 m_\sigma^4} \, \left[2 \partial^2 G^A_{\mu\nu}G^{A, \mu\nu} \, (H^\dagger H)\right].
\eea
and the same $\lambda$ dependence remains and results from the dot product in momenta of the gluons
generating $p_h^2$. Similarly, one can arrange the derivative terms
through mapping Eq.~\eqref{derivativeterms} to the total derivative
\bea
\frac{a \, b \, g_s^2}{2 m_\sigma^4} \, \partial^2 \left[G^A_{\mu\nu}G^{A, \mu\nu} \, (H^\dagger H)\right].
\eea
and $2 \left[\partial_\mu (G^A_{\mu\nu}G^{A, \mu\nu}) \, \partial_\mu (H^\dagger H)\right]$.
This later term again generates the same $\lambda$ dependence through the momentum dot product
for the three point function, with a basis choice that is an alternate to the conventions in
the geoSMEFT, but still projects onto the physical three point amplitude in a consistent fashion.
This indicates that experimental constraints on $\Sigma_\kappa$ do not introduce intrinsic basic
dependence due to the $\lambda$ dependence present in this matching example.

\section{Conclusions}
In this paper we have explored the theory uncertainty on $\sigma(\mathcal G \mathcal G \to h)$ from higher order terms in the SMEFT framework, and how that uncertainty is affected by which pieces of the SMEFT calculation are retained when fitting experimental data. This study is made possible by the calculation of  $\sigma(\mathcal G \mathcal G \to h)$ in Ref.~\cite{Corbett:2021cil}, the first analysis to include both complete $\mathcal O(1/\Lambda^4)$ effects and one loop corrections to $\mathcal O(1/\Lambda^2)$ terms. We explored three ways of splitting the full $\mathcal O(1/\Lambda^4)$, $\mathcal O(1/16\pi^2\Lambda^2)$ result into a subset used for fitting experimental data, and a remainder that defines the uncertainty: i.) fitting experimental data with the linear $\mathcal L^{(6)}$ piece only (in which case, the uncertainty is all of Eq.~\eqref{eq:gghrationumeric} except the terms linear in $C^{(6)}_{HG}$), ii.)  fitting with the linear and quadratic $\mathcal L^{(6)}$ pieces, and iii.) fits including select $\mathcal L^{(8)}$ terms. Defined in this fashion, the theory error is controlled primarily by the dimensionful scale $\Lambda$ and can be combined in quadrature with the experimental uncertainty.

Cases ii.) and iii.) are unconventional as they contain only a subset of higher order results, however they capture physics that case i.) cannot, such as a relative suppression in interference terms relative to $(\mathcal L^{(6)})^2$ terms originating from the fact that $gg \to h$ is a one-loop process in the SM. Incorporating $C^{(8)}_{HG}$ terms into the fit, forming a combination with $C^{(6)}_{HG}$ and $(C^{(6)}_{HG})^2$ we define as $\Sigma_k$, further stabilizes the theory uncertainty when assuming a tree/loop hierarchy of Wilson coefficients. We find that field redefinition ambiguities in cases ii.) and iii.) are small, suppressed by interference with the SM amplitude, and the type of basis dependence $\propto \lambda$, the Higgs quartic, observed in Ref.~\cite{Hays:2020scx,Corbett:2021eux} does not appear to arise.

When extracting numerical results, we explored two different Wilson coefficient schemes,
all coefficients the same, and tree/loop hierarchy. While obviously not exhaustive,
these two schemes span a wide class of UV scenarios; for other setups,
one could repeat the steps here starting with the result in Ref.~\cite{Corbett:2021cil}.

Finally, we wish to stress that the loop nature of $\sigma(\mathcal G \mathcal G \to h)$ in the SM plays a crucial role in the validity of including partial $O(1/\Lambda^4)$ results when comparing with experiment, as it suppresses field redefinition ambiguities on the quadratic term (independent of the Wilson coefficient matching scheme). We strongly stress that our conclusions do not generally apply to the case where a tree level SM amplitude is present to interfere with SMEFT perturbations. When retaining partial $\mathcal O(1/\Lambda^4)$ terms in a projection of experimental results in such a case, numerical ambiguities can be $\mathcal{O}(1)$.

\section{Acknowledgements}
M.T. acknowledges the Villum Fund, project number 00010102.
The work of A.M. was supported in part by the National Science Foundation under Grant Number PHY-1820860 and PHY-2112540.
We thank Tyler Corbett, Chris Hays and Andreas Helset for insightful discussions.

\appendix

\section{Appendix A}\label{appendixa}
An alternative approach to illustrate the effect of higher order terms leading to theory error estimates is
to set $C_{HG}^{(6)}$ to a fixed value, and then illustrate the resulting change in the induced
deviation in $\sigma(\mathcal{G} \mathcal{G} \rightarrow h)$ when the higher order coefficients
are varied over assumed distributions.

\begin{figure}
\begin{minipage}[b]{.45\textwidth}
\includegraphics[width=0.99\textwidth]{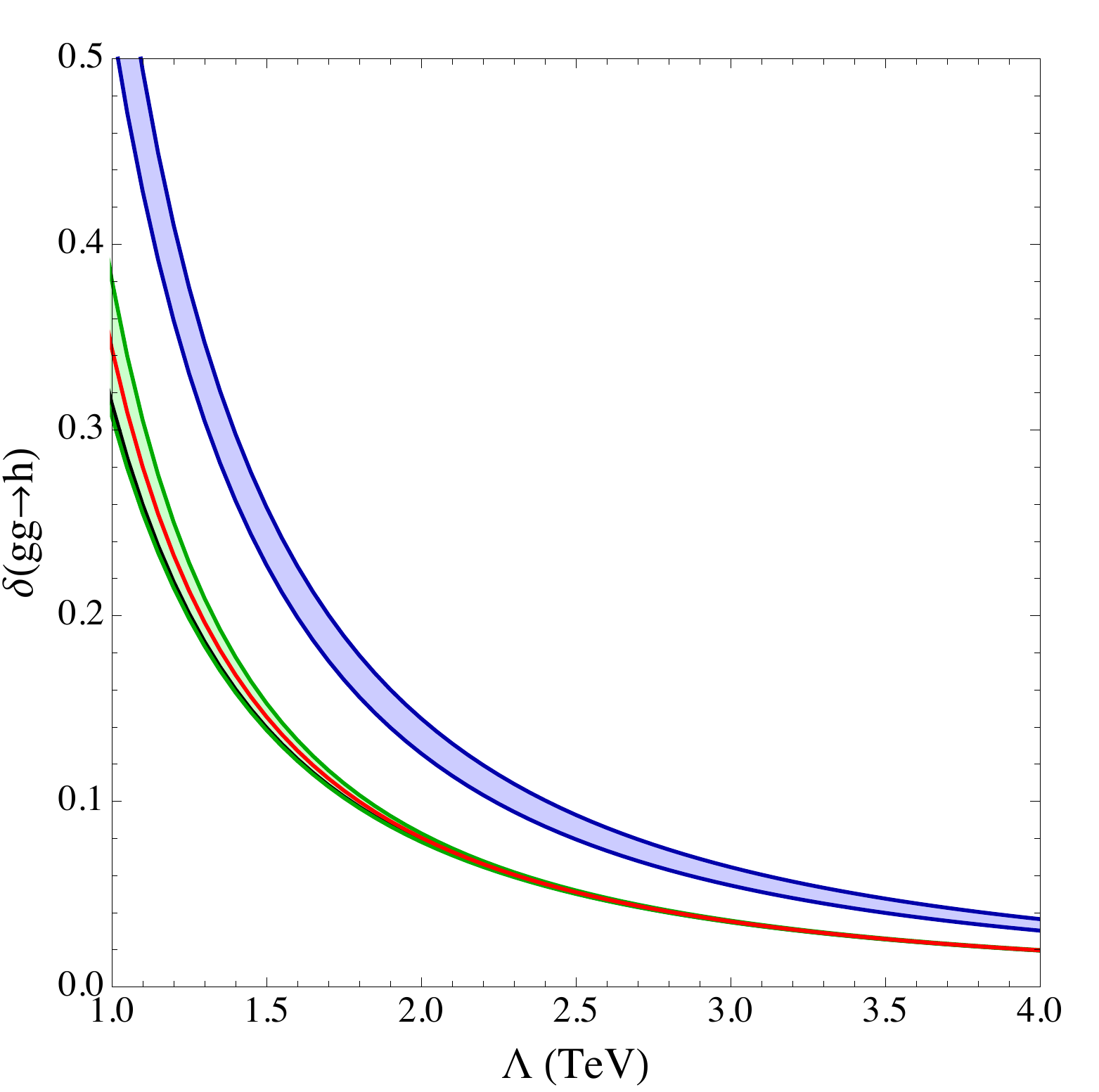}
\caption{Deviation in $\sigma(\mathcal G \mathcal G \to h)$ relative to the SM with $C^{(6)}_{HG} = 0.01$, and all other coefficients sampled according to gaussian distributions with zero mean and width 0.01. The deviation is plotted as a function of $\Lambda$. The black (red) lines correspond to the linear (quadratic) $C^{(6)}_{HG}$ terms, the green band is the $2\sigma$ band that results from 10k samples of the $\mathcal O(1/\Lambda^4)$ corrections, and the blue band is the $2\sigma$ band from 10k samples of the sum of the $\mathcal O(1/\Lambda^4)$ and loop level, $\mathcal O(1/16\pi^2\Lambda^2$) terms.}
\label{fig:all01}
\end{minipage}\\ \qquad
\begin{minipage}[b]{.45\textwidth}
\includegraphics[width=0.99\textwidth]{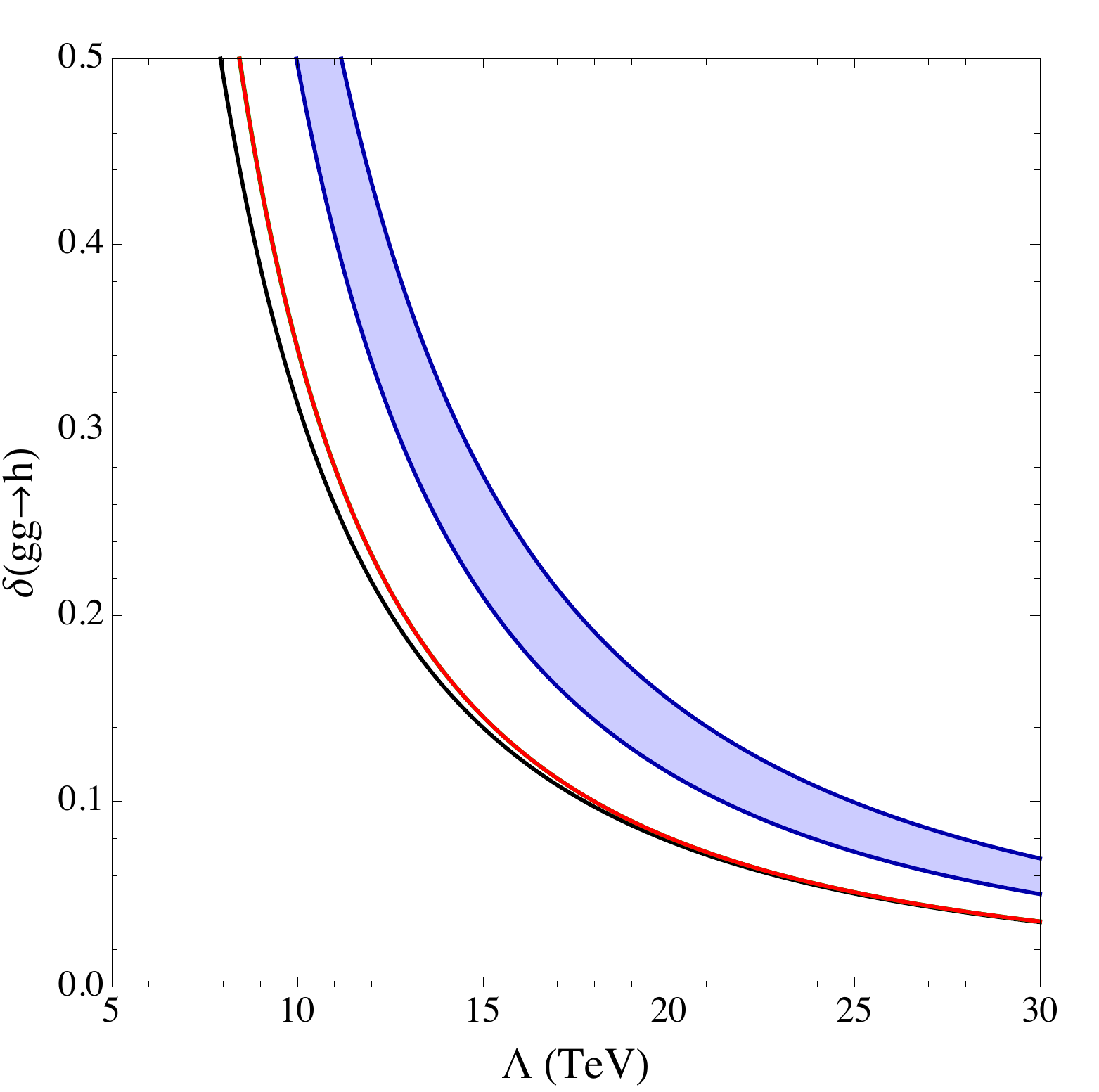}
\caption{Deviation in $\sigma(\mathcal G \mathcal G \to h)$ relative to the SM with $C^{(6)}_{HG} = 1.0$, and all other coefficients sampled according to gaussian distributions with zero mean and width 1.0. The deviation is plotted as a function of $\Lambda$. The black (red) lines correspond to the linear (quadratic) $C^{(6)}_{HG}$ terms, the green band is the $2\sigma$ band that results from 10k samples of the $\mathcal O(1/\Lambda^4)$ corrections, and the blue band is the $2\sigma$ band from 10k samples of the sum of the $\mathcal O(1/\Lambda^4)$ and loop level, $\mathcal O(1/16\pi^2\Lambda^2$) terms.}\label{fig:all10}
\end{minipage}
\end{figure}

\begin{figure}
\includegraphics[width=0.45\textwidth]{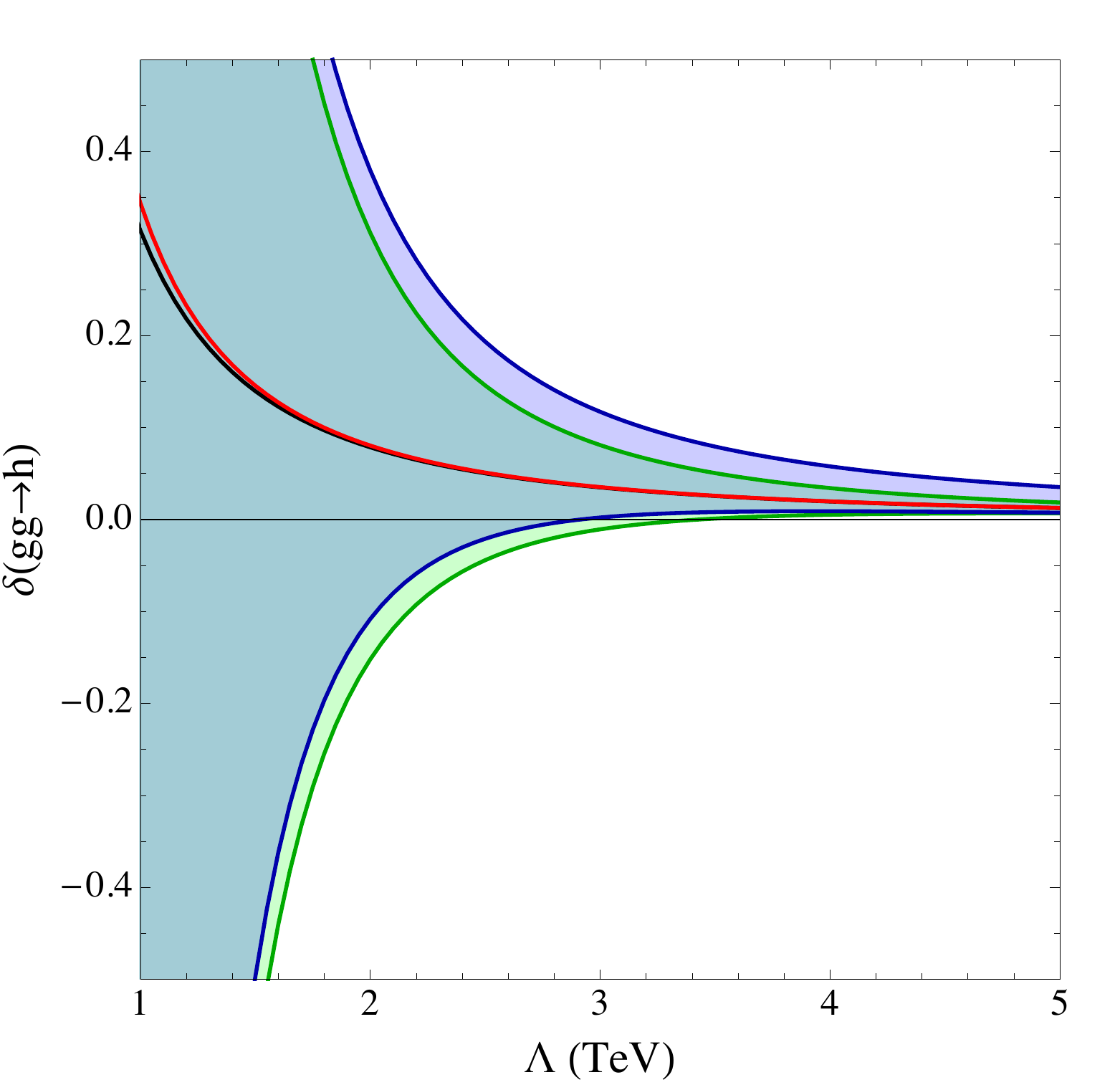}
\caption{Deviation in $\sigma(\mathcal G \mathcal G \to h)$ relative to the SM with $C^{(6)}_{HG} = 0.01$, and all other coefficients sampled according to gaussian distributions with zero mean and width of either 1.0 or .01 depending on whether the corresponding operator is generated at tree or loop level following the classification in Ref.~\cite{Arzt:1994gp, deBlas:2017xtg, Craig:2019wmo}. The deviation is plotted as a function of $\Lambda$, and the color scheme for the lines and bands is the same as in Figs.~\ref{fig:all01}, \ref{fig:all10}.}
\label{fig:treeloop}
\end{figure}

These results are shown in Figs.~\ref{fig:all01},\ref{fig:all10},\ref{fig:treeloop}. In each of the figures, the black and red lines indicate the contribution to Eq.~\eqref{eq:gghrationumeric} from the linear and quadratic $C_{HG}^{(6)}$ terms, respectively. The green band shows the range of values when the $\mathcal O(1/\Lambda^4)$ terms are included, and the blue band shows the range once $\mathcal O(1/\Lambda^4)$ and `loop', $\mathcal O(1/16\pi^2 \Lambda^2)$ terms are included. The range of values correspond to $2\sigma$ values, derived from sampling the coefficients in the higher order ($\mathcal O(1/\Lambda^4)$ or $\mathcal O(1/16\pi^2 \Lambda^2)$) terms $10$k times from gaussian distributions and extracting the standard deviation of the collection. The difference between the figures is the assumptions made on the Wilson coefficients; in Fig.~\ref{fig:all01}, we set $C_{HG}^{(6)} = 0.01$ and sample the higher order terms according to a gaussian with zero mean and width 0.01, in Fig~\ref{fig:all10} we use 1.0 for the value of $C_{HG}^{(6)}$ and the width of the sampling gaussians, and in Fig.~\ref{fig:treeloop} we use a tree/loop scheme -- setting $C_{HG}^{(6)} = 0.01$ and using 1.0/0.01 for the width of the gaussians for operators that fall into the tree/loop category. The horizontal axes of the three figures have been chosen such that the (absolute value of the) deviation in $\sigma(\mathcal G\mathcal G \to h)$ is less than 0.5.

\section{Appendix B}\label{appendixb}
Here we apply the numerical error analysis technique from Sec.~{\bf V} to $\Gamma(h\to\gamma\gamma)$; $\Gamma(h\to\gamma\gamma)$ is also a loop level process in the SM and therefore subject to similar questions as $\sigma(\mathcal G\mathcal G \to h)$ of which SMEFT contributions to keep when projecting experimental results and the impact of higher order terms. The full SMEFT expression to $\mathcal O(v^2_T/16\pi^2\Lambda^2)$, $\mathcal O(v^4_T/\Lambda^4)$ is derived in Ref.~\cite{Corbett:2021cil}
\begin{widetext}
\begin{align}
\frac{\Gamma^{\hat{m}_{W}}_{SMEFT}}{\Gamma^{\hat{m}_{W}}_{\rm SM}}
&\simeq 1 -  788 f^{\hat{m}_W}_1, \\
 &+ 394^2 \, (f^{\hat{m}_W}_1)^2
- 351 \, (\tilde{C}_{HW}^{(6)} - \tilde{C}_{HB}^{(6)})\, f^{\hat{m}_W}_3 + 2228 \, \delta G_F^{(6)} \, f^{\hat{m}_W}_1, \nonumber \\
&+  979 \, \tilde{C}_{HD}^{(6)}(\tilde{C}_{HB}^{(6)} +0.80\, \, \tilde{C}_{HW}^{(6)} -1.02  \, \tilde{C}_{HWB}^{(6)})
-788 \left[ \left(\tilde C_{H\Box}^{(6)} - \frac{\tilde C_{HD}^{(6)}}{4}\right)  \, f^{\hat{m}_W}_1+ f^{\hat{m}_W}_2\right], \nonumber \\
&+2283 \, \tilde{C}_{HWB}^{(6)}(\tilde{C}_{HB}^{(6)} +0.66 \, \, \tilde{C}_{HW}^{(6)} -0.88  \, \tilde{C}_{HWB}^{(6)})
- 1224 \, (f^{\hat{m}_W}_1)^2, \nn
&- 117 \, \tilde{C}_{HB}^{(6)} - 23 \, \tilde{C}_{HW}^{(6)} + \left[51 + 2 \log \left(\frac{\hat{m}_h^2}{\Lambda^2}\right)\right] \, \tilde{C}_{HWB}^{(6)}
+ \left[-0.55 + 3.6 \log \left(\frac{\hat{m}_h^2}{\Lambda^2}\right)\right]\, \tilde{C}_{W}^{(6)}, \nn
&+ \left[27 - 28 \log \left(\frac{\hat{m}_h^2}{\Lambda^2}\right)\right]\, {\rm Re} \, \tilde{C}_{\substack{uB \\ 33}}^{(6)}
+ \left[14 - 15\log \left(\frac{\hat{m}_h^2}{\Lambda^2}\right)\right]\, {\rm Re} \, \tilde{C}_{\substack{uW \\ 33}}^{(6)}
+ 5.5 {\rm} Re \, {\tilde{C}_{\substack{uH \\ 33}}^{(6)}}, \nn
&+ 2 \, \tilde{C}_{H \Box}^{(6)} - \frac{\tilde{C}_{HD}^{(6)}}{2} -3.2 \, \tilde{C}_{HD}^{(6)} -7.5 \, \tilde{C}_{HWB}^{(6)} - 3 \,\sqrt{2} \,\delta G_F^{(6)}.
\label{eq:haa_mW}
\end{align}
\end{widetext}
in the $\hat m_W$ scheme, and
\begin{widetext}
\begin{align}
\frac{\Gamma_{SMEFT}^{\hat{\alpha}_{ew}}}{ \Gamma^{\hat{\alpha}_{ew}}_{\rm SM}} &\simeq
1 -  758 f^{\hat{\alpha}_{ew}}_1, \nn
&+379^2 \, (f^{\hat{\alpha}_{ew}}_1)^2
- 350 \, (\tilde{C}_{HW}^{(6)} - \tilde{C}_{HB}^{(6)})^2 - 1159 \, (f^{\hat{\alpha}_{ew}}_1)^2 \nn
&- 61\,\tilde{C}_{HWB}^{(6)} \, \left(\tilde{C}_{HB}^{(6)} +7.2 \tilde{C}_{HW}^{(6)} -9.2 \tilde{C}_{HWB}^{(6)} \right)
- 13.5\,\tilde{C}_{HD}^{(6)} \, \left(\tilde{C}_{HB}^{(6)} +16 \tilde{C}_{HW}^{(6)} -15 \tilde{C}_{HWB}^{(6)} \right) \nn
&+ 1383\, \delta G_F^{(6)} \, \left(\tilde{C}_{HB}^{(6)}-0.13 \tilde{C}_{HW}^{(6)} -0.15 \tilde{C}_{HWB}^{(6)} \right)
-  758 \left[\left(\tilde C_{H\Box}^{(6)} -\frac{\tilde C_{HD}^{(6)}}{4}\right) f^{\hat{\alpha}_{ew}}_1
+ f^{\hat{\alpha}_{ew}}_2\right],\nn
&- 218 \, \tilde{C}_{HB}^{(6)} +22 \, \tilde{C}_{HW}^{(6)} + \left[-17 + 2.0 \log \left(\frac{\hat{m}_h^2}{\Lambda^2}\right)\right] \, \tilde{C}_{HWB}^{(6)}
+ \left[-0.60 + 3.6 \log \left(\frac{\hat{m}_h^2}{\Lambda^2}\right)\right]\, \tilde{C}_{W}^{(6)}, \nn
&+ \left[26 -27 \log \left(\frac{\hat{m}_h^2}{\Lambda^2}\right)\right]\, {\rm Re} \, \tilde{C}_{\substack{uB \\ 33}}^{(6)}
+ \left[14 - 15\log \left(\frac{\hat{m}_h^2}{\Lambda^2}\right)\right]\, {\rm Re} \, \tilde{C}_{\substack{uW \\ 33}}^{(6)}
+ 5.5 {\rm} Re \, {\tilde{C}_{\substack{uH \\ 33}}^{(6)}}, \nn
&+ 2 \, \tilde{C}_{H \Box}^{(6)} - \frac{\tilde{C}_{HD}^{(6)}}{2}
- \sqrt{2} \,\delta G_F^{(6)}.
\label{eq:haa_alp}
\end{align}
\end{widetext}
in the $\hat \alpha_{ew}$ scheme. Here, $C^{(6+2n)}_{HB}$, $C^{(6+2n)}_{HW}$, $C^{(8)}_{HW,2}$, $C^{(6+2n)}_{HWB}$, $C^{(6)}_{uH}$, $C^{(6)}_{uB}$ and $C^{(6)}_{W}$ are the Wilson coefficients of the following operators:
\begin{align}
\mathcal{Q}_{HB}^{(6+2n)} &= (H^\dag H)^{(1+n)} \, B^{\mu \nu}\, B_{\mu \nu}, \nn
\mathcal{Q}_{HW}^{(6+2n)} &= (H^\dag H)^{(1+n)} \, W_a^{\mu \nu}\, W^a_{\mu \nu}, \nn
\mathcal{Q}_{HW,2}^{(8)} &= (H^\dag \sigma_a H)\,(H^\dag \sigma_b H)\, W_a^{\mu \nu}\, W^b_{\mu \nu}, \nn
\mathcal{Q}_{HWB}^{(6+2n)} &= (H^\dag \sigma_a H)\, (H^\dag H)^{(n)}\, W^a_{\mu \nu}\, B^{\mu \nu}, \nn
\mathcal{Q}_{uH}^{(6)} &= (H^\dag H)(\bar q_r \, u_r \tilde{H}), \nn
\mathcal{Q}_{uB}^{(6)} &= (\bar q_r \, \sigma^{\mu\nu}\, u_r) \tilde{H} B^{\mu\nu}, \nn
\mathcal{Q}_{W}^{(6)} &= \epsilon^{IJK} \, W^{I,\nu}_{\mu} \, W^{J,\rho}_{\nu}  \, W^{K,\mu}_{\rho},
\end{align}
and  $f_i^{\hat m_W} \cong f_i^{\hat \alpha_{ew}}$ are linear combinations of Wilson coefficients:
\begin{align}
f^{\hat{m}_W}_1 &=  \left[\tilde{C}_{HB}^{(6)} +0.29 \, \, \tilde{C}_{HW}^{(6)} -0.54  \, \tilde{C}_{HWB}^{(6)}\right],\\
f^{\hat{m}_W}_2 &=   \left[\tilde{C}_{HB}^{(8)} +0.29 \, \, (\tilde{C}_{HW}^{(8)}+ \tilde{C}_{HW,2}^{(8)}) -0.54  \, \tilde{C}_{HWB}^{(8)}\right],\\
f^{\hat{m}_W}_3 &= \left[\tilde{C}_{HW}^{(6)} - \tilde{C}_{HB}^{(6)} -0.66  \, \tilde{C}_{HWB}^{(6)}\right],
\end{align}

 Following the analysis of $\sigma(\mathcal G\mathcal G \to h)$, we break up the full result for $\Gamma(h \to \gamma\gamma)$ into three cases:
\begin{itemize}
\item[i.)] Retaining only the dimension six interference piece, $\propto f_1$, when comparing with experiment. The loop corrections for $\Gamma(h \to \gamma\gamma)$ are not $\propto f_1$~\cite{Hartmann:2015oia}, so in this case we only keep the tree level interference term.
\item[ii.)] Retaining the interference piece plus $(f_1)^2$ terms, the square of the dimension six piece from i.).
\item[iii.)] Retaining the $f_1$, $(f_1)^2$ and $f_2$ terms.
\end{itemize}
In each case, we associate the remainder of Eq.~\eqref{eq:haa_mW}, \eqref{eq:haa_alp} with the impact from higher order terms and explore its numerical impact using the same two Wilson coefficient matching schemes used in the main text.

We next determine the minimum scale $\Lambda_{\rm min}$ by equating the retained part of $\Gamma(h \to \gamma\gamma)$ to the current uncertainty on $gg \to h \to \gamma\gamma$, $\delta \mu_{gg \to h \to \gamma\gamma} = 0.14$~\cite{Aad_2020} and setting Wilson coefficients according to the matching scheme. Then, for $\Lambda > \Lambda_{\rm min}$, we evaluate the higher order piece $10{\rm k}$ times, evaluating the higher order terms at each step using values drawn from gaussian distributions with width set by the matching scheme. The standard deviation from the collection of higher order term values is shown below in Fig.~\ref{fig:hgamgam_mW},\ref{fig:hgamgam_alpha} as a function of $\Lambda/\Lambda_{\rm min}$ for the various cases, matching, and electroweak input schemes.\footnote{Explicitly, the $\Lambda_{\rm min}$ values for $\Gamma^{\hat m_W}(h \to \gamma\gamma)$ are $\Lambda_{\rm min} = 1.6\, {\rm TeV}$ for all cases when the Wilson coefficients are all 0.01, $\Lambda_{\rm min} = 16\, {\rm TeV}$  for all cases when the Wilson coefficients are all 1.0, and $\Lambda_{\rm min} = 1.6\, {\rm TeV}, 1.6\, {\rm TeV}, 2.5\, {\rm TeV}$ for cases i.), ii.), iii.) respectively in the tree/loop 1.0/0.01 scheme. The $\Lambda_{\rm min}$ values for $\Gamma^{\hat \alpha_{ew}}(h \to \gamma\gamma)$ are essentially the same.}


\begin{figure}[h!]
\begin{minipage}[b]{0.45\textwidth}
\includegraphics[width=0.99\textwidth]{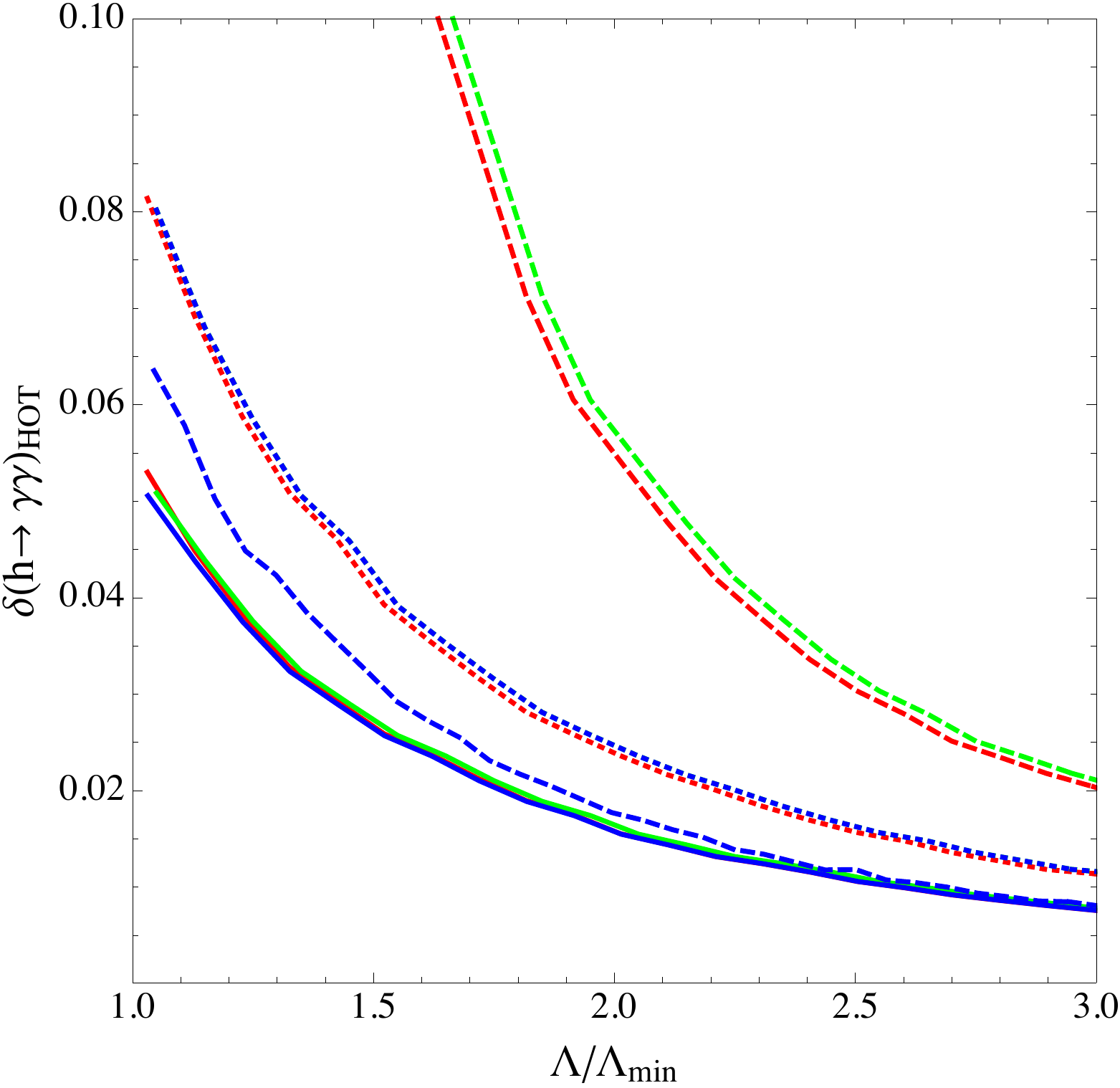}
\caption{Uncertainty on $\Gamma(h \to \gamma\gamma)$ from higher order terms in the $\hat m_W$ scheme. The different colored lines correspond to different ways of breaking up the SMEFT calculation into a piece thats compared to experiment and a higher order correction. The red lines correspond to projecting the data onto Wilson coefficients using the dimension six interference term  ($f_1$) only, the green lines correspond to including the $f_1$ and $(f_1)^2$ pieces, and the blue line corresponds to including $f_1$, $(f_1)^2$ and $f_2$. The solid lines correspond to a coefficient matching scheme where all coefficients are 0.01, the dotted lines correspond to all coefficients equal to 1, and the dashed lines correspond to the tree/loop scheme with values 1.0/0.01.}
\label{fig:hgamgam_mW}
\end{minipage}\\
\begin{minipage}[b]{0.45\textwidth}
\includegraphics[width=0.99\textwidth]{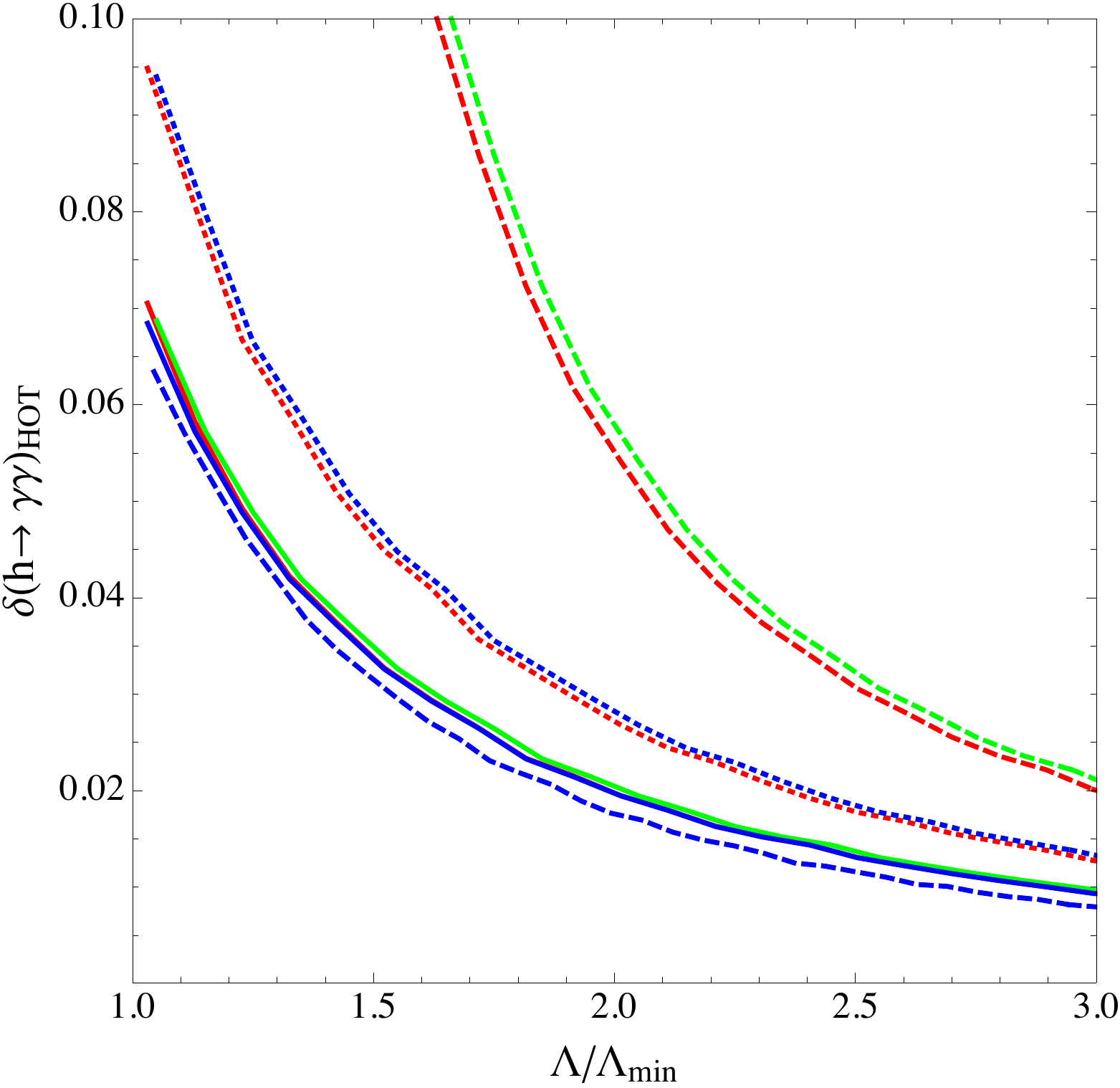}
\caption{Uncertainty on $\Gamma(h \to \gamma\gamma)$ from higher order terms in the $\hat \alpha_{ew}$ scheme. The color and dashing scheme is the same as in Fig.~\ref{fig:hgamgam_mW}}.
\label{fig:hgamgam_alpha}
\end{minipage}
\end{figure}

As was the case in $\sigma(\mathcal G \mathcal G \to h)$, case iii.) is the most robust under the different Wilson coefficient schemes studied here. As was the case for $\sigma(\mathcal G \mathcal G \to h)$, the difference between the curves with all Wilson coefficients equal to 1 and all coefficients equal to 0.01 (when plotted vs. $\Lambda/\Lambda_{\rm min}$) can be traced to the $\log(\Lambda^2)$ terms in $\Gamma(h \to \gamma\gamma)$. Additionally, comparing Figs.~\ref{fig:hgamgam_mW} and \ref{fig:hgamgam_alpha}, one can see there is some dependence on the EW input scheme.

\bibliography{bibliography1}
\end{document}